\documentclass[%
 reprint,
 amsmath,amssymb,
 aps,
]{revtex4-1}
\usepackage{graphicx}
\usepackage{float}
\usepackage{dcolumn}
\usepackage{bm}
\usepackage[utf8]{inputenc}

\begin{document}
\title{Heat-to-Mechanical Energy Conversion in Graphene: Manifestation of Umklapp Enhancement with Strain}
\author{Daryoush Shiri}
\author{Andreas Isacsson}
\email{andreas.isacsson@chalmers.se}
\affiliation{ Department of Physics, Chalmers University of Technology, SE 41296 G\"{o}teborg, Sweden}
\date{\today}

\begin{abstract}
Conversion of heat-flux, from a steady state temperature difference, to mechanical vibration is demonstrated in graphene nanoribbons using direct non-equilibrium molecular dynamics (NEMD). We observe that this effect is independent of the method of imposing the temperature gradient, heat flux as well as imposed boundary conditions. We propose that simply dividing the nanoribbon in long and short sections using a partially immobilized area will lead to excitation of long-wavelength vibrations in the long section of the nanoribbon. This results in simpler architectures for heat-to-vibration converter devices based on graphene or other 2D materials. Furthermore we observe that applying tensile axial strain to nanoribbons, facilitates vibrational instability by reducing the required threshold heat flux or temperature gradient. Finally, we discuss the role played by Umklapp scattering for physical mechanisms behind these observations. 

\begin{description}
\item[PACS numbers]
\end{description}
\end{abstract}
\maketitle
\section{\label{sec:level1} Introduction}

Investigations of methods to engineer and exploit the thermal properties of graphene and other 2D materials for heat transport and thermal management applications are in progress~\cite{Pop_MRS2012,Pop_suspnd13, BaowenLi_random18}. Not only does the high thermal conductivity of graphene make it promising for applications, but also the possibility to observe for instance hydrodynamic heat transport~\cite{Lee_2015}, has sparked a more general interest in non-Fourier heat transport mechanisms. Among these, converting a temperature gradient to mechanical motion is an alternative route to alter the thermal conductivity of graphene, without resorting to complex fabrication methods or processing~\cite{BaowenLi_suspnd11}. 

Here, we are in particular interested in the thermoelastic phenomenon where a steady thermal bias can give rise to spontaneous long-wave length flexural vibrations of a suspended mechanical system -- a phenomenon akin to the thermoacoustic effect. Such thermomechanical instability phenomena gained interest after the first numerical results showing non-Fourier heat transport accompanied by spontaneous excitation of mechanical oscillations in carbon nanotubes (CNTs)~\cite{ZhangNL2012_h2m}. Similar results were later found in simulations of graphene nanoribbons (GNR)~\cite{Zheng2014_h2m, WenJun2016_h2m}, 2D hetero-structures~\cite{XueKun2017_h2m,Liu2014_h2m} and graded nanowires~\cite{YueYang2014_h2m,Liu2015_h2m}. 

Despite the large body of numerical evidence for heat to mechanical energy conversion, the underlying physics of the phenomenon remains poorly understood. A common heuristic argument is that heat transport accompanied by spontaneous mechanical excitation will only take place if the transport via this channel is comparable to that of diffusion or ballistic phonon transport. This argumentation is consistent with numerical observations~\cite{ZhangNL2012_h2m, Zheng2014_h2m, XueKun2017_h2m,YueYang2014_h2m,Liu2015_h2m} of a threshold heat flux needed to induce oscillations, and that increasing the lengths of the systems quenches the mechanical oscillations due to enhanced thermal conductivity which "short circuits" the thermomechanical channel. We here seek to shed further light on the origin of the thermomechanical instability in carbon nanosystems by focusing on graphene nanoribbons.

In accordance with most of the previous studies we use Molecular Dynamics (MD) simulations performed in LAMMPS~\cite{Plimpton_1995}. However, whereas previous studies of CNTs and Graphene nanoribbons employed the AIREBO potential~\cite{AIREBO} along with the M{\"u}ller-Plathe algorithm~\cite{MPlathe} to flux bias the systems, we here observe the same phenomena using a different many-body potential (i.e. Tersoff~\cite{Lindsay_2010}) and non-equilibrium molecular dynamics (NEMD) with thermal biasing with constant temperature difference. This insensitivity to numerical details lends further support to heat-to-vibration conversion being a robust thermoelastic phenomenon which should be readily observable in an experimental setting. 

The above mentioned CNT- and GNR- geometries~\cite{ZhangNL2012_h2m, Zheng2014_h2m, WenJun2016_h2m} are difficult to realize experimentally as they rely on thermal biasing of a freely suspended mid-section of the systems. However, in asymmetric geometries intended for heat rectification~\cite{XueKun2017_h2m,Liu2014_h2m,YueYang2014_h2m,Liu2015_h2m}, thermoelastic vibrational instabilities have been seen in simulations. Still, the proposed 2D-material systems~\cite{XueKun2017_h2m,Liu2014_h2m} require that one can reliably fabricate and contact suspended asymmetric hetero-structures of for instance Graphene-hBN, which may prove challenging from an experimental point of view.  

Here, we show that such hetero-structures are not needed to observe thermoelastic vibrational instabilities. Instead, it suffices to have asymmetric GNRs wherein asymmetry is imposed by a pinned or defected area dividing the ribbon into short and long sections. This obviates the need for complex geometries like hetero-junctions or graded structures. We further find that applying external tensile axial strain to the nanoribbon, lowers the threshold flux required to trigger the heat-to-vibration instability. This highlights the important role played by Umklapp scattering in the heat-to-vibration conversion. Understanding the source of the mechanical vibrations should also shed light on methods for enhancing the heat rectification ratio in nanoribbons~\cite{ShiriIsacsson2017} and in general asymmetric hetero-structure of thermal conductors~\cite{XueKun2017_h2m, YueYang2014_h2m, Liu2015_h2m, Liu2014_h2m}.
\begin{figure}[t]
\centering \includegraphics[width=1\linewidth]{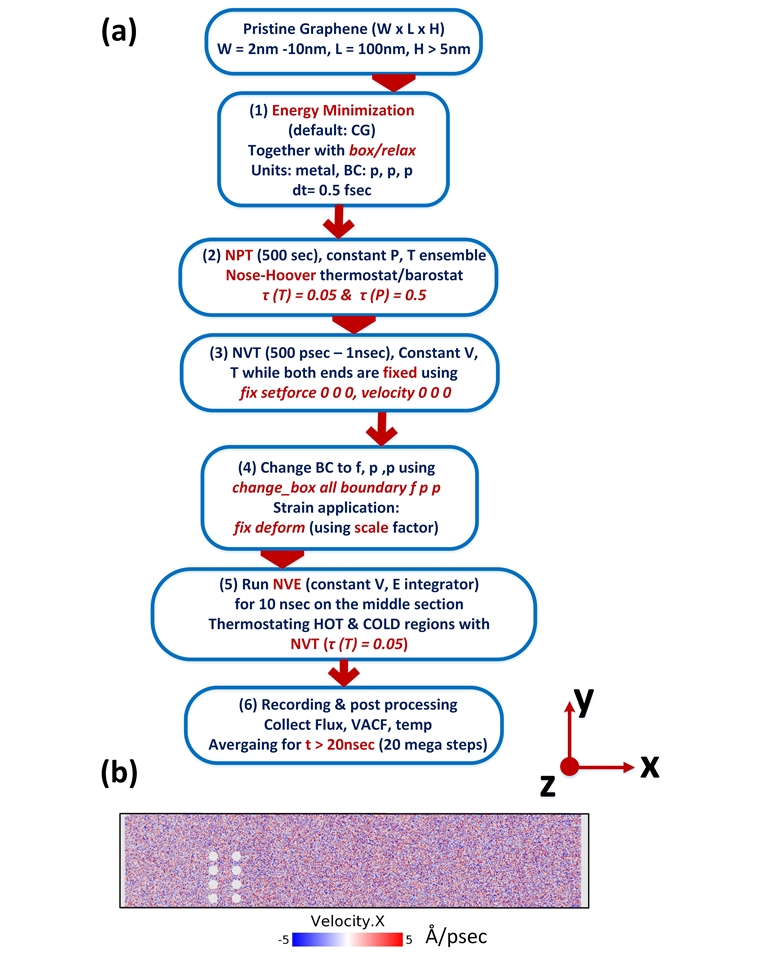}
\caption {(a) Flow of the molecular dynamics simulations with corresponding important commands and settings in LAMMPS. (b) The 20~nm $\times$ 100~nm graphene nanoribbon with zigzag chirality along \emph{x} direction (length). The circular pinned features are 20~Å wide and 10~Å apart. The white ends are fixed and adjacent to these fixed areas are hot (right) and cold (left) terminals.}
\label{fig:methods}
\end{figure}
\section{\label{sec:level1} Methods}
The MD-simulations are performed using the LAMMPS package~\cite{Plimpton_1995}, and Figure~\ref{fig:methods}a sketches all steps with corresponding important commands and settings for each stage of the simulation. The GNRs are of rectangular shape with the long (suspended) edge in the zigzag direction. Their widths and lengths are $20\,$nm and $100\,$nm, along the $y$ and $x$ directions respectively (see Figure~\ref{fig:methods}b). 

The size of the simulation box along the $z$ direction is $\pm50\,$\AA~to avoid interaction of the graphene ribbon with its periodic replica. The graphene nanoribbon is built with the ATOMSK code~\cite{Hirel_2015} starting from a unit cell made of four carbon atoms or using tools like the VMD molecular graphics viewer~\cite{VMD96}. The energy of the ribbon is minimized under periodic boundary conditions in all directions using a Conjugate Gradient algorithm (\emph{box/relax} command in LAMMPS). The Tersoff three-body potential~\cite{Lindsay_2010} which is modified for heat transport in graphene is used. The integration time step is $0.5\,$fs. 

The system is then equilibrated at zero pressure by simulating in the NPT ensemble together with a Nose-Hoover thermostat and barostat for $500\,$ps ($10^6\,$steps). The time constants for temperature and pressure settling are $0.05\,$ps and $0.5\,$ps, respectively. This is followed by a round of NVT ensemble simulations for $500\,$ps~-~$1\,$ns while both ends of the GNR are fixed using zero force and velocity at all time steps. The fixed edges are $10\,$\AA~wide strips of atoms and are from hereon labelled \emph{Hot} and \emph{Cold} in anticipation of the thermal biasing at fixed temperature difference. 

We then change the boundary conditions by fixing the short edges while keeping the long edges either periodic or free. 
Application of tensile strain is achieved by gradually stretching the simulation box along the $x$-direction. The NVE ensemble is used for $10\,$ns for the middle section which is the GNR excluding fixed, Hot and Cold sections. The temperature of Hot and Cold sections are controlled by Nose-Hoover thermostats. After the initial $10\,$ns, heat flux, velocity auto correlation function (VACF) and temperature are window averaged for every $10^4$ steps and the results are recorded for $20\,$ns or more depending on convergence of the temperature profile. 

For visualizing purposes we use the OVITO package~\cite{Stukowski_2010}. The vibrational density of states of atoms within a region is calculated from the velocity auto correlation function (VACF):  
\begin{equation}
\langle v(t)v(0)\rangle= \\
\langle v_{x}(t)v_{x}(0)\rangle+\langle v_{y}(t)v_{y}(0)\rangle+\langle v_{z}(t)v_{z}(0)\rangle. 
\end{equation}
Both $\langle v(t)v(0)\rangle$ as well as individual components $\langle v_{i}(t)v_{i}(0)\rangle$ are recorded in the output file for each time step (e.g. $dt=0.5\,$fs). This allows us to use the Fast Fourier Transform of each component of the VACF to obtain the vibrational density of states (VDOS) spanning frequencies from 0~Hz to around 60~THz. Furthermore the VACF data are used to extract recurrence plots (RP). These provide a useful alternative representation of the information in the VACF (see section~\ref{sec:results}).

\begin{figure}[H]
\includegraphics[width=\linewidth]{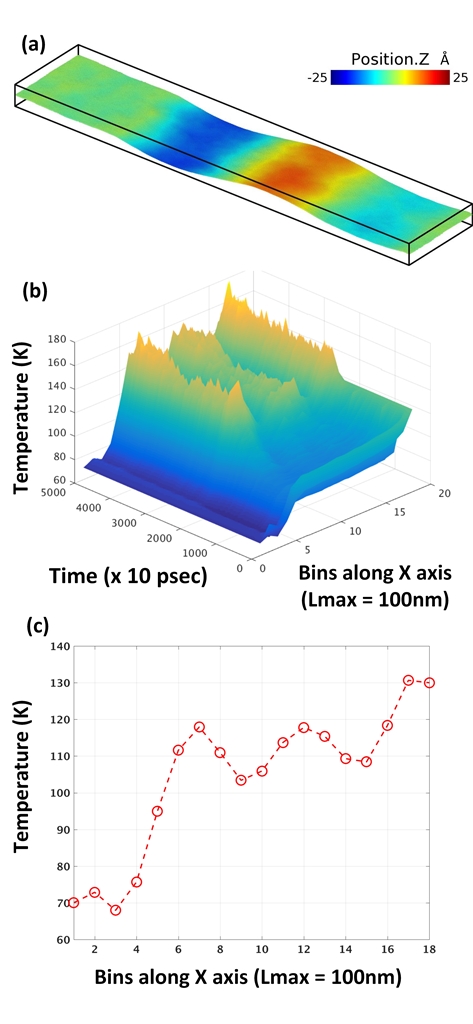}
\caption{(a) Visualization of vibrations of the long section of the nanoribbon when the right terminal is biased at $T_R=$ 130 K. (b) Time averaged temperature profile along the length of the ribbon showing oscillation build-up after $20\,$nsec from the beginning of the of simulation.  Each 1000 sample corresponds to 10~nsec simulation time. (c) The ensemble averaged of temperature profile along the length of the nanoribbon (x). The sudden temperature drop is due to stationary atoms within the forced area which do not contribute to the kinetic energy (temperature) i.e. their velocity is zero}
\label{fig:undulat13070}
\end{figure}

\section{Results\label{sec:results}}
\subsection{Role of Temperature Bias and Polarity}
The $20\,$nm~$\times$~$100$\,nm GNR is divided into a short(left) and a long(right) section using two rows of circular areas with pinned atoms (see Figure ~\ref{fig:methods}b). The hot and cold sections are biased at $T_0 \pm \Delta T$ where $T_0$ ranges from $100\,$K to $400\,$K and $\Delta T$ is varied from $10\,$K to $100\,$K. Figure~\ref{fig:undulat13070}a shows the oscillations of the long section of the nanoribbon for $T_0=100\,$K and $\Delta T = -30\,$K when the long section is biased at higher temperature, i.e. $T_{\rm R}=130\,$K and $T_{\rm L}=70\,$K. 

The time averaged temperature profile in Figure~\ref{fig:undulat13070}b also reveals that when $T_{\rm R} > T_{\rm L}$, the onset of temperature undulations starts after the time step 2000 ($20\,$ns). Figure~\ref{fig:undulat13070}c shows the time averaged temperature profile after $20\,$ns, along the length of the nanoribbon. 

\begin{figure}[H]
\centering \includegraphics[width=\linewidth]{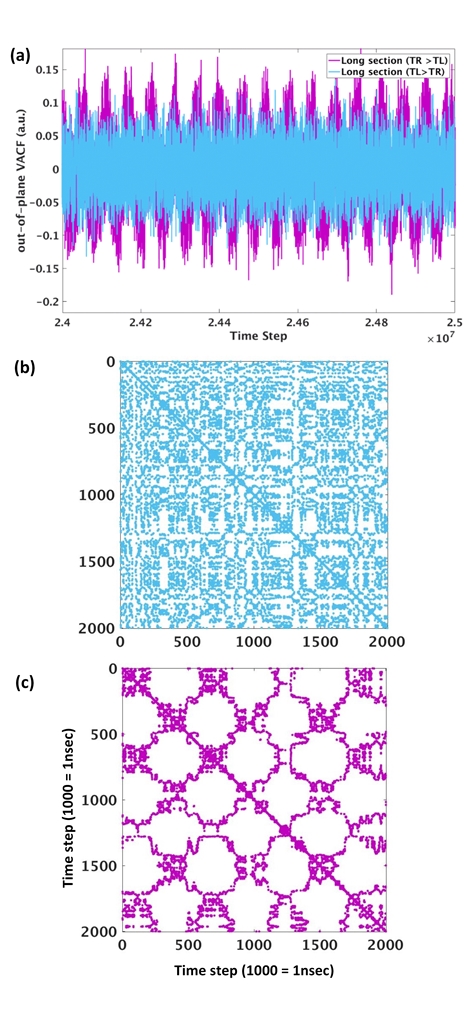}
\caption{(a) Velocity auto-correlation function for out-of-plane $z$ component of velocity for long (right) section of the ribbon at forward and reversed temperature bias of $\Delta T=30\,$K. (b) The recurrence plot corresponding for the forward biased (blue) VACF which shows no evidence of periodicity or harmonic content i.e. it is of white noise character. (c) The same data for the reverse biased (magenta) nanoribbon proving the existence of oscillations.}
\label{fig:recur13070}
\end{figure}

To better illustrate the oscillatory motion, we study the velocity auto-correlation function (VACF) of the long section at two temperature differences having different signs i.e. $\Delta T = \pm 30\,$K. Two snapshots of the VACF for the out-of-plane velocity components ($\langle v_{z}(t)v_{z}(0)\rangle$) are shown in Figure~\ref{fig:recur13070}a, from $24\,$ns to $25\,$ns. The oscillatory behavior for $T_{\rm R} > T_{\rm L}$ as opposed the case when $T_{\rm R} < T_{\rm L}$ is evident. 

An alternative way of visualizing the VACF data, is to generate recurrence plots (RP). These plots, which were proposed by Eckman~\emph{et al.}~\cite{Eckman87}, enable visualization of behavior of a dynamical system and its $m$-dimensional phase space trajectory as a two dimensional matrix. For instance, if $x(i)$ is a point on this trajectory, for $i=1,..., N$ a $N\times~N$ matrix is constructed. If $x(i)$ and $x(j)$ are closer than a predefined threshold (or error) in the trajectory, the element $(i,j)$ in the recurrence plot is 1, otherwise it will be considered 0. Hence each element of the recurrence plot(matrix) i.e. $RP_{i,j}$ is given as:
\begin{equation}
RP_{i,j}= \theta(\epsilon - \parallel \overrightarrow{x}_i - \overrightarrow {x}_j \parallel), 
\overrightarrow{x}_i \in \Re^{m}, i,j = 1, ..., N.
\end{equation}
Here $\theta$ is the Heaviside unit step function, $\epsilon$ is the chosen closeness (recurrence) threshold. The dimension of the dynamical space is denoted by $m$. The recurrence plots of VACF data for $N=2000$ and $\epsilon=0.03$ are shown in Figures~\ref{fig:recur13070}b and c. These plots are visualization of sparse matrices which are obtained by \emph{spy} function in MATLAB. For the forward biased section ($\Delta T = +30\,$K), there is no evidence of periodicity or harmonic content in the RP as it resembles the RP of white noise (Figure~\ref{fig:recur13070}b). On the other hand, quasi harmonic oscillations are signalled by vertical lines to the main diagonal and as multiple lines in parallel. This is clearly seen in RP for the reverse biased ribbon (see Figure~\ref{fig:recur13070}c) when $\Delta T = -30\,$K.

\begin{figure*} [t]
\centering \includegraphics[width=\linewidth]{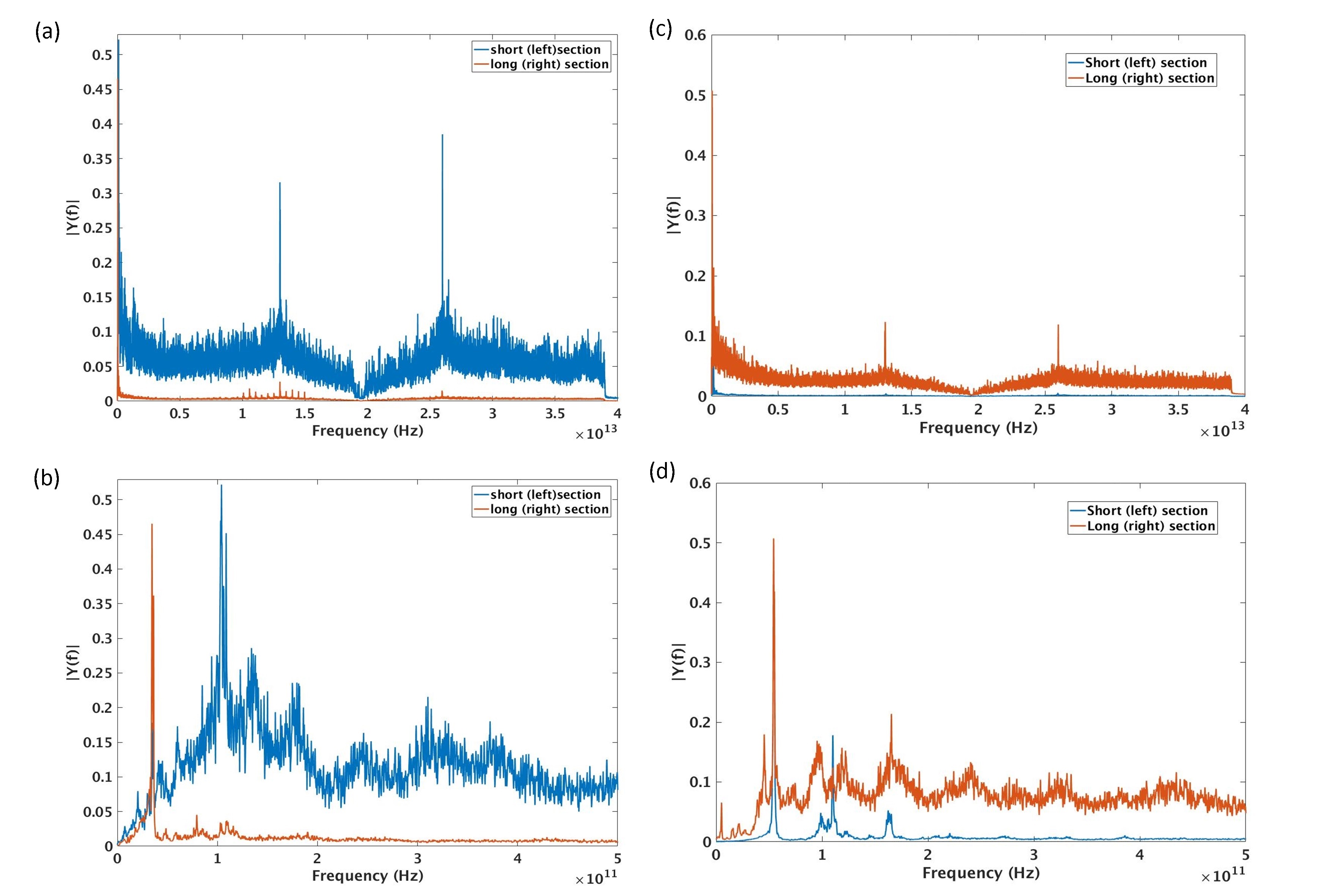}
\caption{Out-of-plane vibrational density of states (VDOS) of short (left) and long (right) sections of the nanoribbon biased at $T_0=100\,$K and $\Delta=\pm 30\,$K. (a) and (b) show the whole spectra and zoomed low frequency part of the spectra as the long section is biased at higher temperature i.e. $T_R=130\,$K. As it is evident the long section has a low frequency component at $34.6\,$GHz. (c) and (d) are the same data for the case of forward bias i.e. $T_R=70\,$K. Here the long part has suppressed modes (no mechanical vibration) compared with the short section.}
\label{fig:vdosd30RLLR}
\end{figure*}

To quantify these low frequency modes, the vibrational density of states (VDOS) is extracted through Fourier transforming the VACF. Figures~\ref{fig:vdosd30RLLR}a and b show the full spectrum and the low frequency part of VDOS for $T_{\rm R}= 130\,$K and $T_{\rm L}=70\,$K for out of plane modes respectively. The $13\,$THz and $26\,$THz modes and the gap between them are clearly visible in the spectrum from the short section (see Figure~\ref{fig:vdosd30RLLR}a). The long section on the other hand has two features: (i) A frequency comb around $13\,$THz mode and (ii) a low frequency mode at $34.6\,$GHz.

The frequency comb is an artifact of using periodic boundary conditions in the $y$-direction. As the width of the nanoribbon is $20\,$nm, having flexural modes with propagation velocity of $10^3\,$m/s explains the existence of such combs with spacing of around $0.5\,$THz. The short section also has a larger frequency component at $104.1\,$GHz (see Figure~\ref{fig:vdosd30RLLR}b).

Reversing the temperature bias, firstly reverses the relative strength of the vibrational modes within the high frequency section of the spectrum (see Figure~\ref{fig:vdosd30RLLR}c), and, secondly, induces low frequency modes in the short (left) section of the nanoribbon (see Figure~\ref{fig:vdosd30RLLR}d). By increasing the magnitude of the temperature difference to $\Delta T = -40\,$K and $\Delta T = -50\,$K, the low frequency modes of the long section move to $60\,$GHz and $\sim125\,$GHz, respectively, i.e., higher frequency (higher energy) out-of-plane modes are excited.  

\begin{figure}[t]
\centering \includegraphics[width=\linewidth]{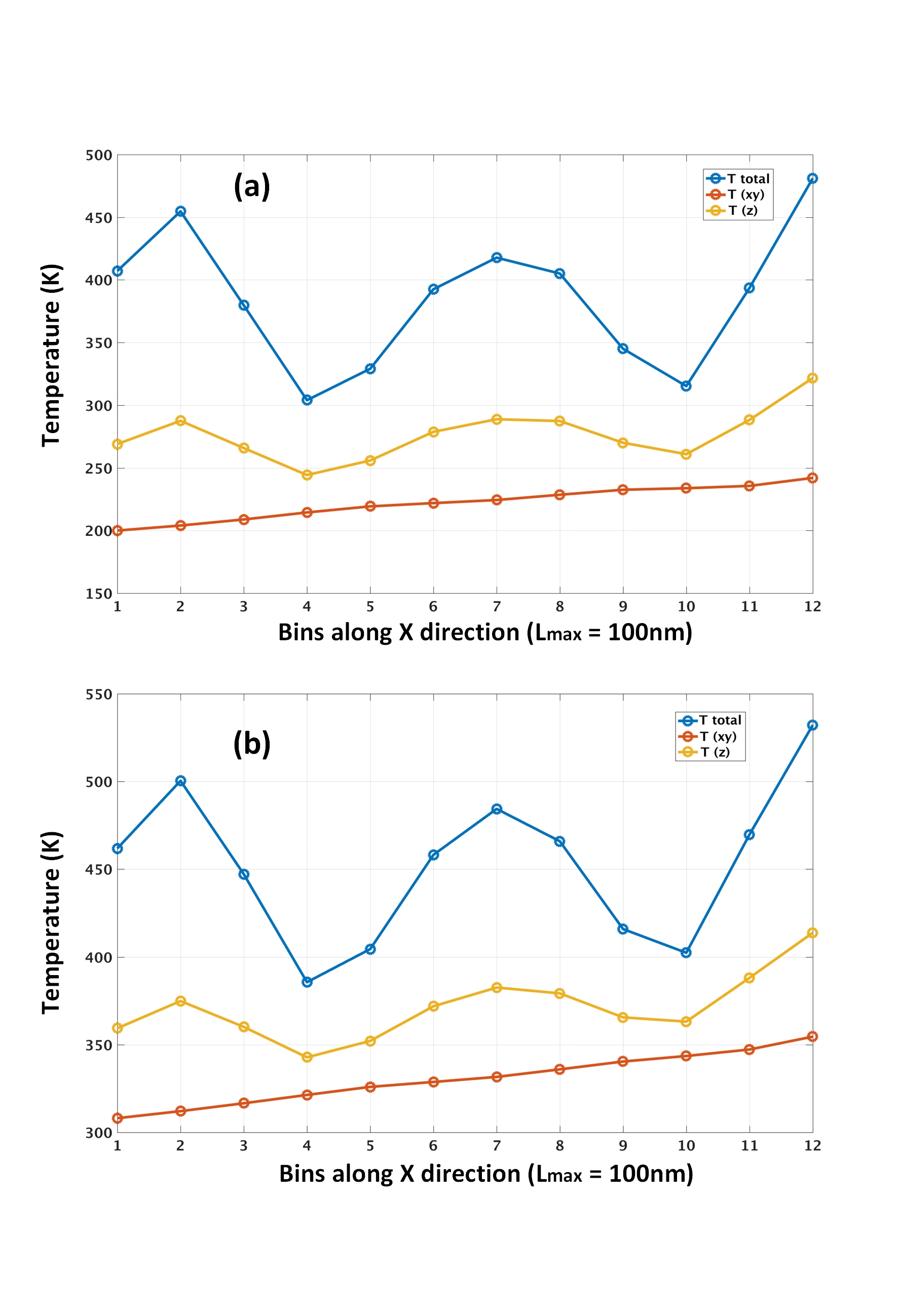}
\caption{Decomposition of energy into random and mechanical parts normalized as temperature. The in-plane contribution is due to random kinetic energy which results in linear gradient i.e. Fourier Type of heat conduction. The mechanical parts is seen as an undulating profile contributing to the total apparent temperature profile. For (a) and (b), the equilibrium temperature is at $T_0=200\,$K and $300\,$K, respectively, with $\Delta T=-50\,$K for both cases.}
\label{fig:TxTyTz200300}
\end{figure}

\subsection{Role of Equilibrium Temperature}
By increasing the equilibrium temperature from $T_0=100\,$K to $200\,$K and $300\,$K respectively, the low frequency out-of-plane vibration still exists. The modes manifest themselves as low frequency modes within the VACF of the long section for negative temperature gradients. For $T_0 =200\,$K and $T_0 = 300\,$K with $\Delta T= -100\,$K, the low frequency modes are around $44\,$GHz and $42\,$GHz, respectively. 

As it was seen before, the undulation in temperature is the sign of transfer of energy by mechanical vibration. To prove that we subtract the random part of the kinetic energy (its equivalent temperature) from the total temperature profile as it is also explained in \cite{ZhangNL2012_h2m}. In agreement with previous observations of real temperature profile in CNT~\cite{ZhangNL2012_h2m} and pristine graphene~\cite{WenJun2016_h2m}, the remaining profile is linear along the length of the long section of the nanoribbon right before the interface. The temperature drop at the interface is due to the fixed (pinned) atoms which contribute no energy in the temperature computation (as it was shown in Figure~\ref{fig:undulat13070}c). The result is shown in Figure \ref{fig:TxTyTz200300} for $T_0 =200\,$K and $300\,$K  with $\Delta T=-50\,$K. The proof of heat-to-mechanical conversion is exemplified here by subtracting the out-of-plane mode energies from the total kinetic energy. The remaining part which is mainly due to random vibrations of atoms creates the linear Fourier type profile showing that the temperature undulation is due to out-of-plane vibrations.  

\begin{figure}[t]
\centering \includegraphics[width=\linewidth]{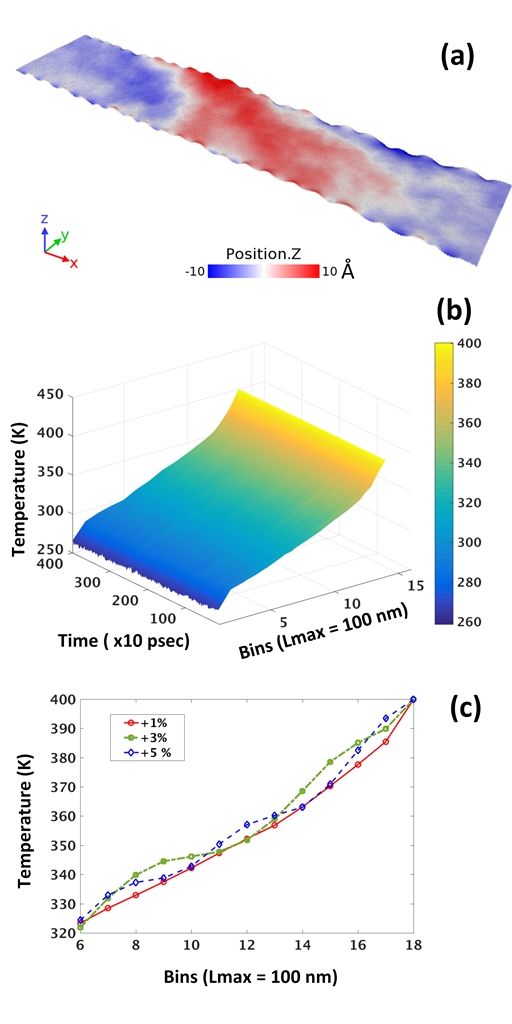}
\caption{(a) Formation of edge modes for the nanoribbon with free boundary condition along $y$-direction. (b) Temperature profile for the long section of the nanoribbon showing no evidence of undulation. (c) Effect of strain in initiating the temperature undulation. The plot is the time averaged of the temperature profile.}
\label{fig:T300_novib}
\end{figure}

\subsection{Role of Mechanical Strain and Free (non-periodic) boundaries}

While periodic boundary conditions are convenient for simulation, we consider in this section free (non-periodic) boundaries along $y$ direction. For the equilibration temperature of $T_0 =300\,$K with $\Delta T=-100\,$K, there is no more evidence of out-of-plane vibration except the vibrations at the edge of the nanoribbon as shown in Figure~\ref{fig:T300_novib}a. Moreover the temperature profile is linear and lacks any undulations (see Figure~\ref{fig:T300_novib}b). However, by increasing the strain from $+1\,\%$ to $+5\,\%$, the thermomechanical vibrational instability is recovered (see Figure~\ref{fig:T300_novib}c). This is in contrast to the claims of references~\cite{ZhangNL2012_h2m,WenJun2016_h2m} that using free boundary condition leads to no qualitative change of results.   
\begin{figure}[t]
\centering \includegraphics[width=\linewidth]{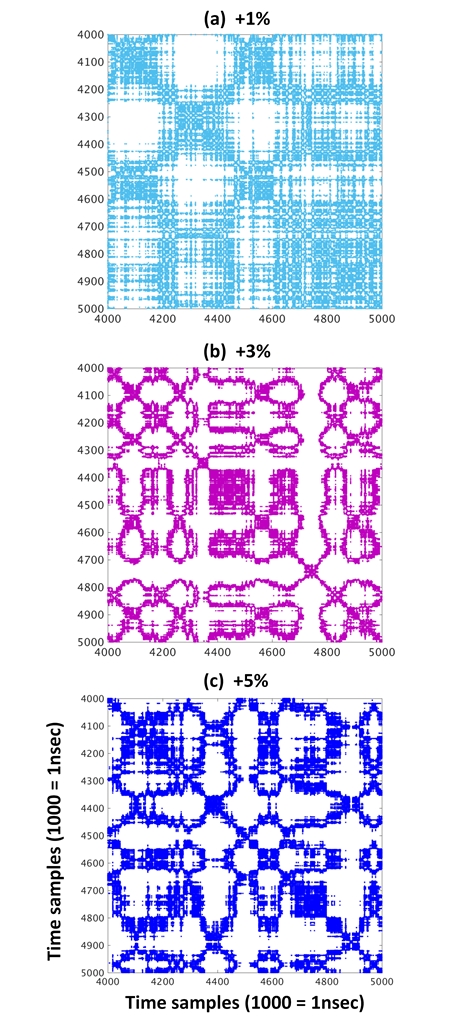}
\caption{Recurrence plots for out-of-plane VACF data from $20\,$ns to $25\,$ns for (a) $+1\,\%$, (b) $+3\,\%$, and (c) $+5\,\%$ strain. $T_0 = 300\,$K with $\Delta T=100\,$K and $1000$ points of data of RP are shown here. Recovery of periodic features are evident by increasing the tensile strain.}
\label{fig:Recur_strain}
\end{figure}

The recurrence plots of VACF data for out-of-plane mode reveal how the harmonic content of the VACF is enhanced by increasing the strain from $+1\,\%$ to $+3\,\%$ (see Figure~\ref{fig:Recur_strain}). To understand this we first consider the role of the changed boundary conditions (from periodic to free) in suppressing the vibrational instability. Adding free boundaries introduces edge scattering of phonons. This additional scattering channel effectively reduces the rate of Umklapp scattering necessary to populate the low frequency out-of-plane modes.

\begin{figure}[t]
\centering \includegraphics[width=\linewidth]{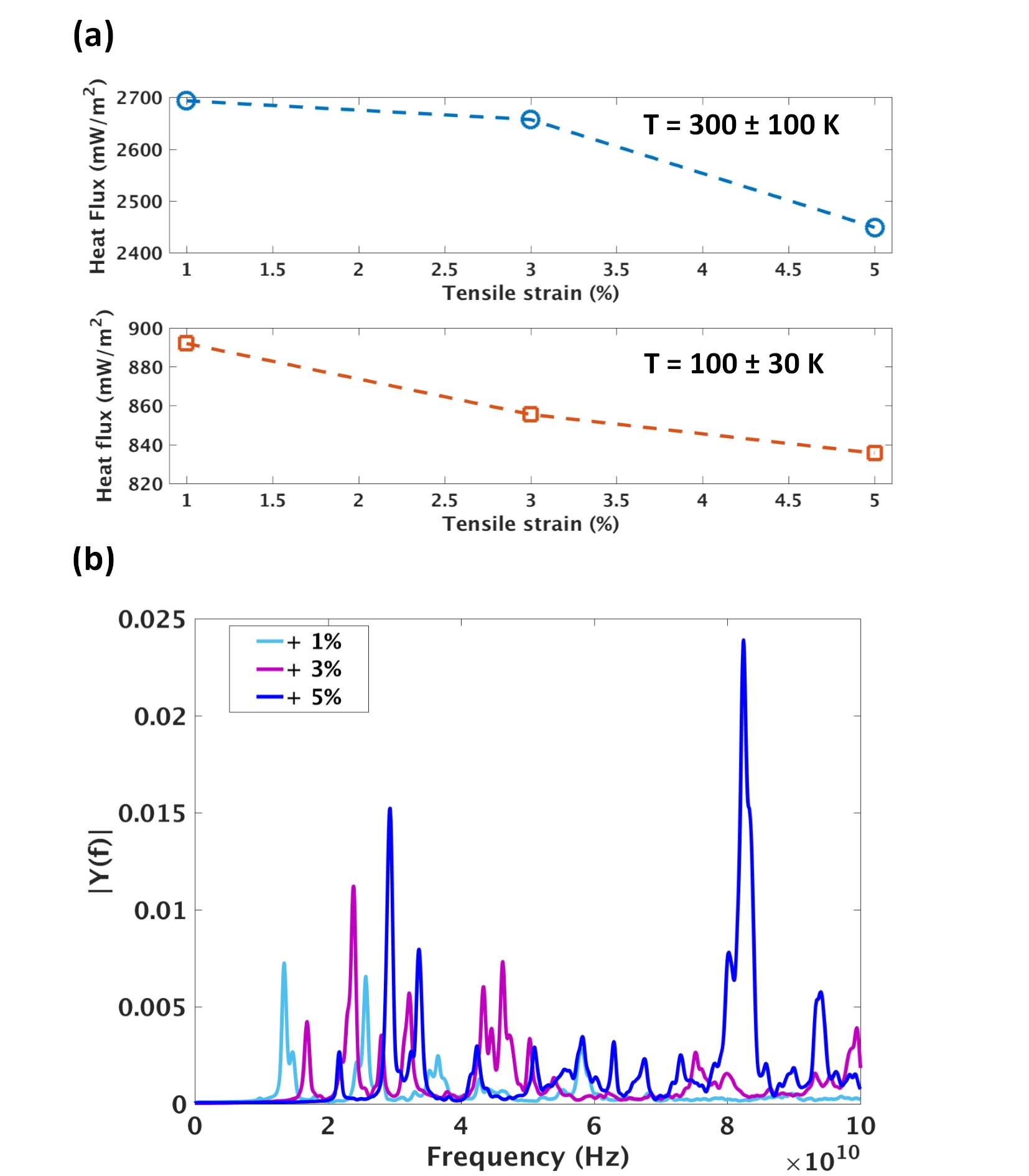}
\caption{(a) Heat flux required for initiation of oscillation plotted for $T_0 = 300~K$ (top panel) and $T_0 = 100~K$ (bottom panel) versus strain. 
(b) Low frequency section of VDOS for out-of-plane mode at 300~K for three strain values. Stiffening of the main acoustic mode is evident by increase of frequency form 0.134~THz to 0.2158 THz.}
\label{fig:VDOSflux_strain}
\end{figure}

However, application of tensile strain increases Umklapp scattering rates~\cite{Bhowmick06}. The corresponding phonon life times $\tau$ scale as 
\begin{equation}
\tau \propto \frac{1}{T} \epsilon^{-\gamma},
\end{equation}
where $\gamma$ is a positive potential-dependent exponent as defined in \cite{Bhowmick06}, $\epsilon$ is tensile strain and $T$ is the temperature. The above equation suggests it is possible to initiate the heat-to-vibration conversion by increasing the tensile strain as it enhances the rate of Umklapp processes i.e. high energy to low energy phonon conversion. This also manifests itself as the reduction of required heat flux from $(900-2700)~ mW/cm^{2}$ to $(840-2450)~mW/cm^{2}$ by increasing the tensile from $+1\%$ to $+3\%$. Figure~\ref{fig:VDOSflux_strain}a confirms reduction of heat flux by increasing the tensile strain for both high and low temperature cases with $T_0 =100\,$K and $300\,$K with $\Delta T=-50\,$K and $-30\,$K, respectively.
Alternatively it could be said that applying tensile strain lowers the frequency of optical phonon modes ($G^{\pm}$) and splits the degeneracy between these modes~\cite{KentoTada17,PNAS_Huang09}. As a result less value of heat flux is required to initiate a scattering (high energy to low energy conversion process) in the space of available secondary momentum/energy states within a given Brillouin Zone (BZ).  

In addition to increase of Umklapp scattering rate, increasing the tensile strain at the same time leads to stiffening of acoustic branches~\cite{KentoTada17}. This is also visible by looking at low frequency section of out-of-plane VACF in Figure \ref{fig:VDOSflux_strain}.b wherein the frequency of the main mode is moving from 0.134~THz, 0.1687~THz, and 0.2158~THz for $+1\%$, $+3\%$ and $+5\%$, respectively. Increase of acoustic phonon frequency with applied tensile strain ($\epsilon$) is understood using (see Ref.~\cite{Gurevitch}):
\begin{equation}
\omega = \omega_{0}~\epsilon^{-a},
\end{equation}
where $a$ is the Gr\"uneisen parameter of the crystalline solid defined as $a=-\partial {\ln\,\omega} /\partial {\ln\,\epsilon}$.

\subsection{Geometry and design simplification\label{sec:simpl}}
Motivated by our previous proposal for a heat rectifier~\cite{ShiriIsacsson2017}, here a more simplified design of a heat to mechanical energy converter is proposed. As mentioned before, a few hetero-junctions of graphene and other 2D materials were proposed in order to benefit from their difference in phonon spectrum which lead to mechanical vibration, heat rectification as well as negative differential thermal conductivity (NDTC)~\cite{XueKun2017_h2m,Liu2014_h2m}. However nano-structuring these hetero-junctions free from defects and other fabrication artifacts is not straightforward. Similarly lithographically patterning a small array of fixed sections in order to asymmetrically dividing a nanoribbon into a short/long or thin/thick sections does not seem straightforward. We propose a strip of forced section on the nanoribbon which as we saw before is also possible to be forced and pinned by electric field~\cite{ShiriIsacsson2017}.

\begin{figure}[t]
\centering \includegraphics[width=\linewidth]{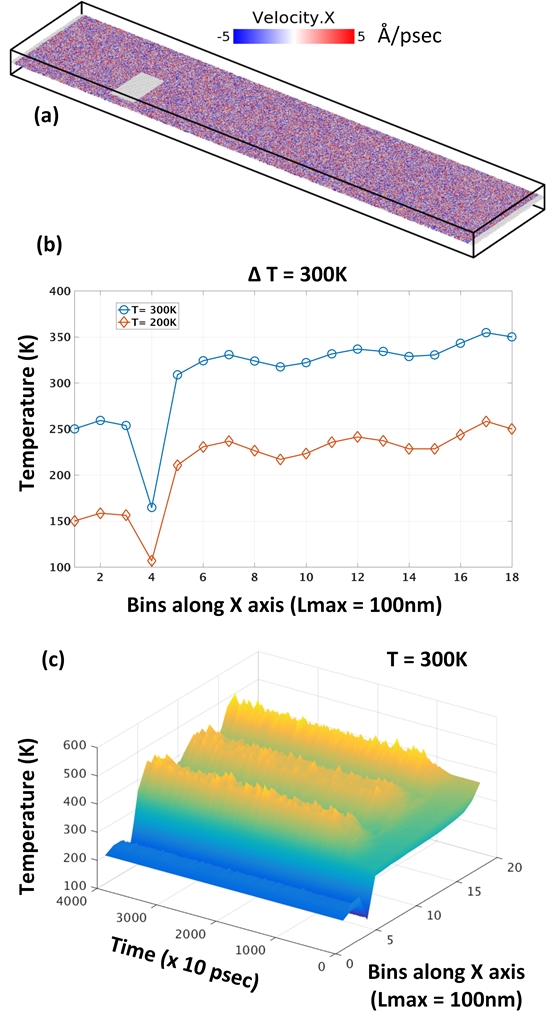}
\caption{(a) Graphene nanoribbon wherein the array of pinned circles is replaced by a simpler pattern which is implementable by applying force. (b) The time averaged temperature profile for 200~K and 300~K with $\Delta T$ of -50 K. (c) The existence of heat to vibration conversion as $\Delta T$ is increased to -100 K for $T_0 = 300~K$.}
\label{fig:patchdata}
\end{figure}

In order to test the idea, instead of an array of pinned circles, a simple rectangular patch of forced atoms is made between both ends of the ribbon (Figure \ref{fig:patchdata}a). The same simulation scenario was repeated for higher base temperatures i.e. $T_0=200\,$K and 300~K with $\Delta T =\pm100\,$K in order to show the suitability of this simpler design for the heat-to-vibration conversion. Figure~\ref{fig:patchdata}a shows the bird's eye view of the graphene nanoribbon with simplified forced area. The time averaged temperature profiles for both $200\,$K and $300\,$K with $\Delta T=-50\,$K are given in Figure~\ref{fig:patchdata}b. This proves that even in this simple structure, merely asymmetric division of the nanoribbon is what is sufficient for heat to vibration conversion. Increasing the temperature gradient to $\Delta T=100\,$K as shown in Figure~\ref{fig:patchdata}c, enhances the amplitude of oscillation as it was expected. 

To probe the role of geometry, the length and width of the nanoribbon is now doubled each time. If only the length is doubled (L = 200~nm), we notice that for $T_0=$200~K and $\Delta T=$ 50~K there is no oscillation as opposed to the original length (L = 100~nm). This means that increasing the length has increased the non-Umklapp (normal) scattering processes which generate fewer low frequency phonons. Interestingly, by doubling the width of the nanoribbon (W = 40~nm) and keeping the forced area and lengths the same as before, the heat to vibration conversion reappears. The only difference is the frequency comb spacing, which is now around 0.25~THz. 

\section{Conclusion}
Using direct Non-equilibrium Molecular Dynamics implemented in LAMMPS we observed conversion of heat to mechanical vibration in an symmetrically-divided graphene nanoribbon at room temperature. The asymmetric division is possible using nano patterned features or simply a section of graphene forced by electric or mechanical means as proposed before \cite{ShiriIsacsson2017}. This implementation is more feasible than the hetero-junctions proposed so far and theoretical set-ups in which one of the hot/cold terminals is afloat in the middle of device. 

Using Tersoff many-body potential, direct NEMD i.e. using temperature gradient as input and recording the flux as output, having different boundary conditions and different kind of asymmetry(interfacing), prove that heat-to-mechanical energy conversion is not an artifact of simulation/method and/or boundary conditions. 

We showed that the mechanism behind heat-to-vibration conversion is the enhancement of the Umklapp process rate. Increasing the heat flux or equally, temperature gradient, excites high momentum and high energy phonons at the boundary of BZ. The higher the number of these phonons as well as the higher number of branches due to zone folding, increases the available space for phonon scattering. 

The increase in Umklapp rate creates more and more low frequency phonons. Quenching the energy conversion effect as a result of (a) non-periodic boundary condition (edge scatterings), (b) increase of length and (c) decrease of temperature gradients are all proving that Umklapp process is the main necessary ingredient of such phenomenon of promising applications. As increasing the length, reduces the number of available modes due to fewer number of zone folding, less frequent Umklapp leads to fewer low frequency modes and this quenches the oscillation. This explains why in previous works by increasing the length, more heat flux was required to compensate the scarce scattering space for phonons. Increase of Umklapp scattering rate by increasing the temperature gradient as well as tensile strain was evidenced by by our simulation results.

\bibliography{Heat_2_Mechanical_Shiri_Isacsson.bib}
\end{document}